\documentclass[epj]{svjour}
\usepackage{graphics}
\usepackage{graphicx}
\usepackage[usenames,dvipsnames]{color}
\begin{document}
\title{A high phase-space density mixture of $^{87}$Rb and $^{133}$Cs:\\ towards ultracold heteronuclear molecules.}

\author{H. W. Cho\thanks{These authors contributed equally to this work.}, D. J. McCarron$^{\rm{a}}$, D. L. Jenkin, M. P. K\"oppinger, and S. L. Cornish.
}
\offprints{s.l.cornish@durham.ac.uk}
\institute{Department of Physics, Durham University, Rochester Building, Science Laboratories, South Road, Durham, DH1 3LE, United Kingdom}
\date{Received: date / Revised version: date}
\abstract{
We report the production of a high phase-space density mixture of $^{87}$Rb and $^{133}$Cs atoms in a levitated crossed optical dipole trap as the first step towards the creation of ultracold RbCs molecules via magneto-association. We present a simple and robust experimental setup designed for the sympathetic cooling of $^{133}$Cs via interspecies elastic collisions with $^{87}$Rb. Working with the $|F=1, m_F=+1 \rangle$ and the $|3, +3 \rangle$ states of $^{87}$Rb and $^{133}$Cs respectively, we measure a high interspecies three-body inelastic collision rate $\sim 10^{-25}-10^{-26}~\rm{cm}^{6}\rm{s}^{-1}$ which hinders the sympathetic cooling. Nevertheless by careful tailoring of the evaporation we can produce phase-space densities near quantum degeneracy for both species simultaneously. In addition we report the observation of an interspecies Feshbach resonance at 181.7(5)~G and demonstrate the creation of Cs$_{2}$ molecules via magneto-association on the 4g(4) resonance at 19.8~G. These results represent important steps towards the creation of ultracold RbCs molecules in our apparatus.
} 
\authorrunning{H. W. Cho \emph{et al.}}
\titlerunning{A high phase-space density mixture of $^{87}$Rb and $^{133}$Cs: towards ultracold heteronuclear molecules.}

\maketitle

\section{\label{chap:Intro}Introduction}

Ultracold molecular quantum systems offer many new and exciting directions of scientific research \cite{Carr2009}. Theoretical proposals for these systems range from precision metrology \cite{Zelevinsky2008} and ultracold chemistry \cite{Krems2008} to simulations of many-body quantum systems \cite{Pupillo2009}. Within this field polar molecules are currently a subject of significant interest. Such molecules possess permanent electric dipole moments which give rise to  anisotropic, long range dipole-dipole interactions. These interactions differ greatly from the isotropic, short-range contact interaction commonly encountered in quantum degenerate atomic gases. The orientation of these dipoles can be controlled by applying an external electric field. This control, in combination with control of the trapping geometry, enables the interactions within the quantum system to be tuned with exquisite sensitivity and may be used to suppress inelastic collisions \cite{Avdeenkov2006}. When loaded onto an optical lattice the long-range interaction of these dipoles leads to a rich spectrum of quantum phases \cite{Capogrosso-Sansone2010} and offers potential applications for quantum information processing \cite{DeMille2002} and simulation \cite{Micheli2006}.

Realising an ultracold quantum gas of polar molecules is challenging, in part due to the complex internal structure that makes molecules interesting. This complexity is due to additional rotational and vibrational degrees of freedom and renders standard laser cooling techniques ineffective for the majority of molecules. Experimental approaches towards the creation of a quantum degenerate gas of polar molecules have mostly followed two routes.

The first approach involves the direct cooling of ground-state polar molecules. This has led to the development of many new experimental techniques \cite{Carr2009},  including Stark deceleration \cite{Bethlem1999} and buffer gas cooling \cite{Weinstein1998}.  Typically these methods have only attained final temperatures in the millikelvin regime and phase-space densities far from degeneracy. Further advances in direct cooling will probably require the development of sympathetic cooling \cite{Modugno2001} or laser cooling \cite{Stuhl2008,Shuman2010}.

The second approach uses the powerful laser cooling techniques applicable to atoms and starts with a high phase-space density atomic gas. Atom pairs are then associated into ground state molecules via an appropriate photoassociation scheme in which the molecular binding energy is removed by a photon  \cite{Jones2006}. The challenge of this indirect approach is to find an efficient scheme in which the population is transferred into a single target molecular state without any heating of the sample. In recent years there have been a large number of successes using the indirect approach and several schemes have produced molecules in their rovibrational ground state \cite{Sage2005,Deiglmayr2008,Viteau2008,Lang2008}. Of particular significance are two experiments that have produced ground state molecular samples close to quantum degeneracy in KRb \cite{Ni2008} and Cs$_{2}$ \cite{Danzl2008,Danzl2010}. These experiments follow a two step scheme in which weakly bound molecules are first made via magneto-association using a Feshbach resonance \cite{Kohler2006}. The resulting Feshbach molecules are then transferred into the rovibrational ground state using stimulated Raman adiabatic passage (STIRAP) \cite{Bergmann1998}. The efficiency of this transfer can be enhanced using the magnetic field to tune the character of the weakly bound Feshbach molecule. Although this approach is limited to molecules whose constituent atoms can be laser cooled, near 100~\% conversion efficiency is possible with little heating of the gas.

In this context an atomic mixture of $^{87}$Rb and $^{133}$Cs offers several key advantages over other atomic mixtures. Crucially, when in the rovibrational ground state, the RbCs molecule is stable against atom exchange reactions as the process RbCs + RbCs $\rightarrow$ Rb$_2$ + Cs$_2$ is endothermic \cite{Zuchowski2010}. There have also been a large number of interspecies Feshbach resonances already identified in this mixture \cite{Pilch2009}, offering numerous resonances to explore magneto-association. In addition there are a number of other benefits including complimentary scattering properties. The $^{133}$Cs scattering length has a rich Feshbach structure with a large background value while the $^{87}$Rb scattering length is essentially independent of magnetic field. This allows the scattering length of one component of the mixture to be tuned without affecting the other. When in the $|F=1,m_F=\pm1 \rangle$ and  $|3,\pm3 \rangle$ states for $^{87}$Rb and $^{133}$Cs respectively the magnetic moment-to-mass ratio differs by less than $2~\%$ for bias fields less than $\sim90~$G. This allows both species to be levitated against gravity using the same magnetic field gradient and ensures excellent spatial overlap between the two clouds. The mass ratio of these species allows rethermalisation to be achieved after few interspecies elastic collisions \cite{Delannoy2001}. Both of these features are important prerequisites for efficient sympathetic cooling.

Here we report our progress towards the production of ultracold polar RbCs molecules. The success of the association scheme described above hinges on three important steps: 1. The production of a high phase-space density sample of the constituent atoms. 2. The magneto-association of weakly bound molecules via a Feshbach resonance. 3. The optical transfer of the molecules into the rovibrational ground state via STIRAP. The majority of this paper describes how the first step is achieved. Our experimental setup and cooling procedure are presented in sections \ref{chap:ExperimentalOverview} and \ref{chap:HighPSD} respectively. We then detail our progress towards the magneto-association of RbCs Feshbach molecules. In Section \ref{chap:FeshbachRes} we present the observation of a suitable interspecies Feshbach resonance and in Section \ref{chap:Molecules} we demonstrate the creation of Cs$_{2}$ molecules to test the experimental protocol. A summary of the results of the paper and an outlook towards the production of RbCs molecules in the rovibrational ground state is then given in the final section.

\textcolor{red}{ }

\section{\label{chap:ExperimentalOverview}Experimental Overview}

\begin{figure}[h t]
\centering
\includegraphics[angle=0, trim = 13mm 75mm 80mm 13mm, clip, scale=0.451]{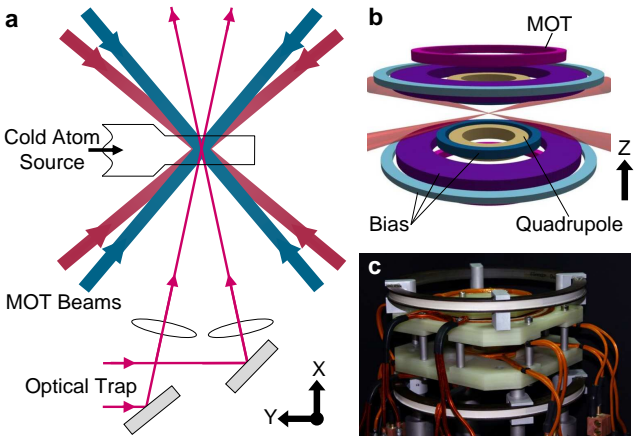}
\caption{Schematic of the experimental setup. (a) Optical beam layout showing the paths of the MOT beams and the beams used to make the crossed dipole trap, intersecting in the centre of a UHV glass cell. The cell is located at the centre of the magnetic coil assembly shown in (b) and (c). The assembly contains two anti-Helmholtz coils pairs for the operation of the MOT and quadrupole trap. A further three coils pairs in Helmholtz configuration can be used to create a uniform bias field at the position of the atoms. }
\label{fig:ExpSetup}
\end{figure}

The experimental setup is designed to simultaneously trap $^{87}$Rb and $^{133}$Cs in their absolute internal ground states. The heart of the setup consists of a levitated crossed dipole trap  \cite{Jenkin2011} located at the centre of an ultrahigh vacuum (UHV) glass cell. This trapping potential consists of two intersecting laser beams, a magnetic field gradient from an anti-Helmholtz coil pair and a uniform bias field from a Helmholtz coil pair. The magnetic gradient is set to $31.1$~G~cm$^{-1}$ to cancel gravity and a variable bias field is applied. The bias field is typically between 21-24~G for evaporation, but can be increased to over 1150~G when searching for Feshbach resonances. The bias field offsets the field zero position to below the crossed dipole trap such that high field seeking states are levitated. Fig. \ref{fig:ExpSetup} presents a  schematic overview of the experimental geometry. The optical contribution to the potential is provided by a 30~W IPG fibre laser with a wavelength of 1550~nm. Two 6~W beams are focussed using lenses with $200~$mm focal lengths to waists of $\sim60~\mu$m. This setup creates a trapping potential 90~$\mu$K (125~$\mu$K) deep for $^{87}$Rb ($^{133}$Cs) and therefore permits the sympathetic cooling of $^{133}$Cs via $^{87}$Rb in the dipole trap. Magnetic field coils are mounted in a coil assembly centred around the glass cell, see Fig. \ref{fig:ExpSetup}. The coils are wound from square cross-section copper tubing and are water cooled. This enables currents exceeding 400~A to be applied with no adverse heating effects. These high currents allow tight magnetic confinement and large bias fields to be applied.

This setup is highly versatile and can create several useful trapping potentials. By blocking one dipole beam a single beam dipole trap can be employed with axial confinement provided by the magnetic quadrupole field \cite{Lin2009}, although this potential can only trap low-field seeking states. In contrast, a two beam crossed dipole trap can confine all spin-states and allows the magnetic field to become a free parameter which can be used to tune intra- and interspecies interactions. To align the crossed dipole trap the position of each individual beam was optimised in the combined magnetic and optical potential configuration \cite{Jenkin2011,Lin2009}. Both beams were positioned $\sim80~\mu$m below the field zero of the magnetic quadrupole potential.

Ultracold atomic mixtures of $^{87}$Rb and $^{133}$Cs are collected in an UHV glass cell using a two-species 6 beam magneto-optical trap (MOT). This is fed from a two-species pyramid MOT. The atoms are prepared into the $^{87}$Rb $|1, -1 \rangle$ and the $^{133}$Cs $|3, -3 \rangle$ states and loaded into a magnetic quadrupole trap at $40$~G~cm$^{-1}$.  Up to $6(1)\times10^8$ $^{87}$Rb and $3(1)\times10^8$ $^{133}$Cs atoms can loaded into the magnetic trap \cite{Harris08}. Active control of the $^{133}$Cs atom number is achieved by monitoring the MOT fluorescence signal with feedback to the $^{133}$Cs repump light level. With this technique the $^{133}$Cs atom number loaded into the magnetic trap can be accurately and reproducibly varied between $2.5(5)\times10^5$ and $3(1)\times10^8$ atoms. Absorption imaging is used to probe the atoms and can analyse both $^{87}$Rb and $^{133}$Cs in one cycle of the experiment.

\section{\label{chap:HighPSD}Cooling to High Phase-Space Density}

\begin{figure}[h t]
\centering
\includegraphics[angle=0, trim = 13mm 13mm 17mm 13mm, clip, scale=0.323]{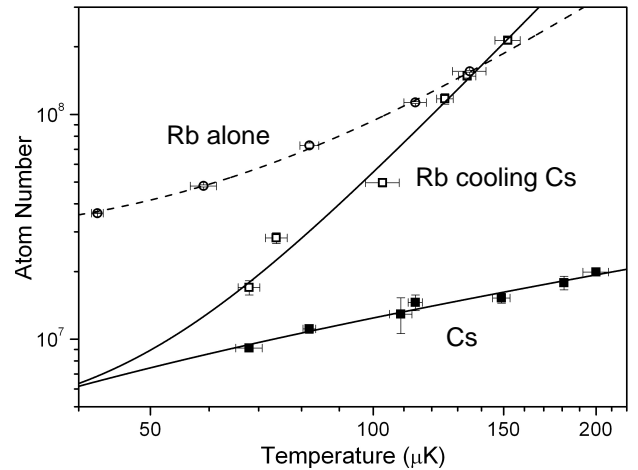}
\caption{Sympathetic cooling of $^{133}$Cs by $^{87}$Rb in the magnetic trap. Forced RF evaporation is used to cool $^{87}$Rb while interspecies elastic collisions cool $^{133}$Cs sympathetically. Open (closed) symbols show data for $^{87}$Rb ($^{133}$Cs). Circular symbols show the evaporative cooling of $^{87}$Rb alone. Square symbols show the two species sympathetic cooling case.}
\label{fig:cooling}
\end{figure}

In order to load the levitated crossed dipole trap the mixture is first precooled with forced RF evaporation in the quadrupole trap. To increase the elastic collision rate the magnetic trap is adiabatically compressed to $187$~G~cm$^{-1}$. In the magnetic quadrupole trap the trap depth set by the RF frequency is three times deeper for $^{133}$Cs than for $^{87}$Rb. This allows the selective RF evaporation of $^{87}$Rb while interspecies elastic collisions sympathetically cool $^{133}$Cs. This sympathetic cooling in the magnetic trap is presented in Fig. \ref{fig:cooling}. The evaporation efficiency is defined as $\gamma = -\frac{log(D_{0}/D)}{log(N_{0}/N)}$, here $N_{0}$ and $D_{0}$ are the initial atom number and phase-space density respectively, $N$ and $D$ are the final number and phase-space density respectively. In the magnetic trap the evaporation efficiency for $^{87}$Rb alone is 2.5(2) (circular data points in Fig. \ref{fig:cooling}). For the two-species sympathetic cooling case the cooling efficiencies are 1.7(1) and 10(1) for $^{87}$Rb and $^{133}$Cs respectively (square data points in Fig. \ref{fig:cooling}). As expected the efficiency of the $^{87}$Rb cooling has decreased during the sympathetic cooling compared to the single species case due to the additional heat load from the $^{133}$Cs atoms. In contrast, the $^{133}$Cs cooling efficiency is very high in accord with the successful demonstration of sympathetic cooling. During the sympathetic cooling the $^{133}$Cs temperature is always observed to be in excellent agreement with the $^{87}$Rb temperature. This indicates rapid cross thermalisation and therefore suggests a large interspecies scattering length. This observation is consistent with the latest coupled channels calculations based upon the large number of published Feshbach resonances \cite{HutsonPC}, and the observation of a bound state very close to threshold using magnetic field modulation spectroscopy \cite{Lercher2011}.

Once the mixture is cooled below $\sim70~\mu$K Majorana spin-flip losses start to limit the efficiency of any further cooling in the magnetic trap. At this point the crossed dipole trap is loaded by elastic collisions as the magnetic field gradient is reduced to 29.0~G~cm$^{-1}$, slightly below the 31.1~G~cm$^{-1}$ required to exactly levitate the atoms. Further benefits of interspecies elastic collisions are observed when loading the dipole trap. By carefully selecting the composition of the mixture and the final frequency of the RF applied during the load, up to 50~\% of the $^{133}$Cs can be transferred into the dipole trap. This is double the number of atoms loaded compared to the $^{133}$Cs alone case. We believe this enhancement results from elastic collisions with the $^{87}$Rb atoms and gives further evidence for a large interspecies elastic cross-section. Once in the levitated crossed dipole trap the phase-space densities are $1.3(1)\times10^{-3}$ and $5.8(5)\times10^{-4}$ for $^{87}$Rb and $^{133}$Cs respectively. This represents a significant increase from the final phase-space densities in the quadrupole trap. During the loading the adiabatic expansion of the quadrupole potential cools the cloud while elastic collisions transfer atoms into the much tighter harmonic optical potential \cite{Lin2009}. The remaining hotter atoms are lost from the quadrupole trap, removing energy in the process. This effect is analogous to the use of a dimple potential \cite{Weber2003}. Immediately after loading the dipole trap the atomic spins of the mixture are flipped into $|1, +1 \rangle$ for $^{87}$Rb and $|3, +3 \rangle$ for $^{133}$Cs via adiabatic rapid passage as a 22.4~G bias field is turned on in 18~ms \cite{Jenkin2011}. This transfer to the absolute ground states means that inelastic two-body losses are avoided. At 22.4~G the ratio of elastic to three-body inelastic collisions in $^{133}$Cs is favourable for the production of Bose-Einstein condensates \cite{Weber2003}.

\begin{figure}[h t]
\centering
\includegraphics[angle=0, trim = 13mm 13mm 13mm 12mm, clip, scale=0.323]{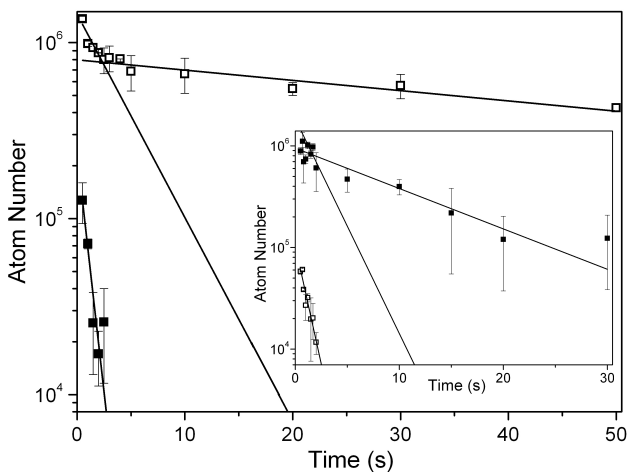}
\caption{Interspecies three-body loss in the crossed dipole trap. Open (closed) symbols show data for $^{87}$Rb ($^{133}$Cs). The data are fit with a double exponential function to determine trap lifetimes. In the main plot $^{133}$Cs is the minority species and the lifetimes are $^{133}$Cs $\tau = 0.8(1)$~s, $^{87}$Rb $\tau_{1} = 4(1)$~s, and $^{87}$Rb $\tau_{2} = 70(10)$~s. In the inset $^{87}$Rb is the minority species and the lifetimes are $^{87}$Rb~$\tau = 0.9(2)$~s, $^{133}$Cs $\tau_{1} = 2(1)$~s, $^{133}$Cs $\tau_{2}= 10(3)$~s.}
\label{fig:3bodyloss}
\end{figure}

After loading the crossed dipole trap the mixture rapidly comes into thermal equilibrium with the potential at $k_{\rm{B}}T\approx {U}_{\rm{Rb}}/10$, where ${U}_{\rm{Rb}}$ is the dipole trap depth for $^{87}$Rb. The peak atomic densities are typically $\approx 5\times10^{13}$~cm$^{-3}$ for $^{87}$Rb and $\approx 5\times10^{12}$~cm$^{-3}$ for $^{133}$Cs. When both $^{87}$Rb and $^{133}$Cs are present in the trap at these high densities very strong interspecies inelastic losses are observed, see Fig. \ref{fig:3bodyloss}. These losses scale with the mean squared density ($\langle n^{2} \rangle$) and so are most prevalent at the trap centre causing the coldest atoms to be lost. This `anti-evaporation' causes both atom loss and heating and is a major hurdle when trying to achieve a high phase-space density mixture. During this process the heating leads to additional losses through evaporation and contaminates our measurement of the three-body loss rate. Due to this we do not perform a full analysis using two coupled differential rate equations to model the evolving densities \cite{Barontini2009}. Instead we perform a naive calculation that uses the mean atomic density for each species, assumes the trap lifetime is due exclusively to three-body losses and does not account for any evaporation from the trap. This makes the calculated loss rate coefficients upper limits but is useful in obtaining an order of magnitude estimate. To test this method the single species $^{133}$Cs case was investigated (inset of Fig. \ref{fig:3bodyloss} for t$\ge 5~$s) and gave a three-body loss rate coefficient of $\sim10^{-27}~\rm{cm}^{6}~\rm{s}^{-1}$, in agreement with previous measurements \cite{Weber2003}. It is this three-body loss in $^{133}$Cs that results in the trap lifetime for t$\ge 5~$s being a factor of seven shorter than that for $^{87}$Rb. The naive calculations for the two species cases used the minority species trap lifetime and yielded interspecies three-body loss rate coefficients of $\sim 10^{-25}-10^{-26}~\rm{cm}^{6}\rm{s}^{-1}$. These unusually high loss rate coefficients imply a large interspecies scattering length and are consistent with our observations and other measurements \cite{Pilch2009}.

\begin{figure}[h t]
\centering
\includegraphics[angle=0, trim = 13mm 13mm 13mm 13mm, clip, scale=0.328]{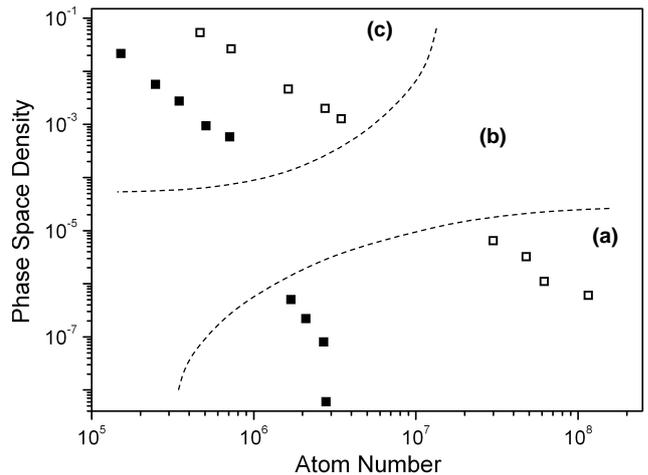}
\caption{Trajectory to high phase-space density for a mixture of $^{87}$Rb and $^{133}$Cs. Open (closed) symbols show data for $^{87}$Rb ($^{133}$Cs). The trajectory is divided into three sections: (a) RF evaporation and sympathetic cooling in the magnetic trap, (b) loading the levitated crossed dipole trap, and (c) evaporative cooling in the levitated crossed dipole trap.}
\label{fig:Trajectory}
\end{figure}

To combat the interspecies three-body loss in the dipole trap fast evaporative cooling is performed. By reducing both beam powers to 120~mW in 1.0~s the $^{87}$Rb trap depth is reduced to $2~\mu$K. This stage not only cools the mixture but also significantly relaxes the trap and results in the atomic density in the trap decreasing by a factor of $\sim$3. Despite this decrease in density and the associated decrease in the elastic collision rate the ratio of elastic to inelastic collisions is increased, improving the conditions for efficient evaporative cooling. As a result the lifetime of the minority species against interspecies three-body collisions increases by one order of magnitude. The evaporation efficiencies for this ramp in the dipole trap are 1.9(1) and 2.4(1) for $^{87}$Rb and $^{133}$Cs respectively. After this cut the mixture consists of $4.7(1)\times10^{5}$ $^{87}$Rb atoms and $1.5(1)\times10^{5}$ $^{133}$Cs atoms at $0.32(1)~\mu$K, corresponding to phase-space densities of $0.053(1)$ and $0.021(1)$ for $^{87}$Rb and $^{133}$Cs respectively. This is the highest reported phase-space density mixture of $^{87}$Rb and $^{133}$Cs produced in a single potential via sympathetic cooling and surpasses previous results working with magnetically trappable states \cite{Anderlini2005,Haas2007}. The final trap lifetime of the minority species ($^{133}$Cs) is 8(2)~s. The trajectory to a high phase-space density mixture of $^{87}$Rb and $^{133}$Cs is summarised in Fig. \ref{fig:Trajectory}. At present it is not possible to extend the evaporative cooling in the dipole trap to lower temperatures due to technical noise on the dipole trap at low powers\footnote{Since writing the manuscript improvements in the dipole trap intensity servo have led to the production of degenerate mixtures of $^{87}$Rb and $^{133}$Cs using this setup \cite{McCarron2011}.}.

\section{\label{chap:FeshbachRes}Interspecies Feshbach Resonance}

\begin{figure} [h t]
\centering
\includegraphics[angle=0, trim = 13mm 13mm 13mm 12mm, clip, scale=0.333]{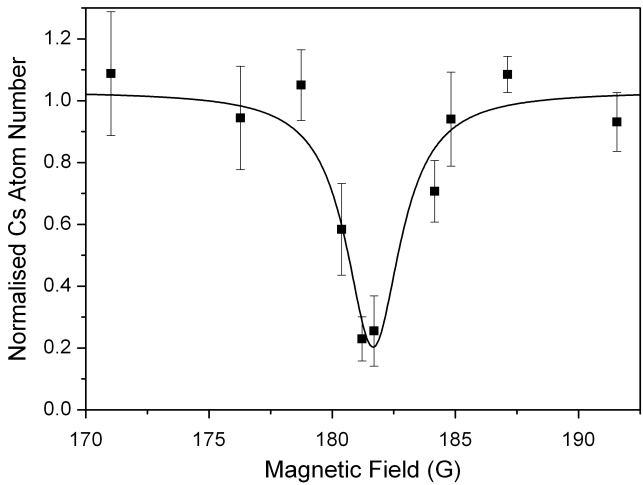}
\caption{Observation of an interspecies $^{87}$Rb$^{133}$Cs Feshbach resonance using loss spectroscopy. The $|1, +1 \rangle$ and $|3, +3 \rangle$ states were used for $^{87}$Rb and $^{133}$Cs respectively. The $^{133}$Cs (minority species) atom number after a 5~s hold is normalised to the corresponding mean number recorded off resonance. A Lorentzian fit is applied to the data to yield a resonance position of $B_{0} =181.7(5)$~G and a width $\approx3$~G.}
\label{fig:RbCs}
\end{figure}

Once a high phase-space density $^{87}$Rb $^{133}$Cs mixture has been made in the levitated crossed dipole trap the search for interspecies Feshbach resonances can begin. Feshbach resonances are most easily detected through an enhancement of trap loss \cite{Chin2010}. To increase the sensitivity of heteronuclear Feshbach spectroscopy a significant imbalance between the two-species atom numbers is useful \cite{Wille2008}. Here the majority species acts as a collisional bath for the minority species which is used as a probe. If the atom number imbalance is large the probe species will be significantly depleted when a resonance is encountered, leaving the majority species largely unchanged. In this experiment the lowest internal energy states are used, this results in trap loss being due to three-body collisions only. To perform sensitive Feshbach spectroscopy a high atomic density for both species is essential as interspecies collisions govern the loss rates from the trap. For the data presented the densities are $1.6(1)\times10^{12}$~cm$^{-3}$ and $3.1(4)\times10^{11}$~cm$^{-3}$ for $^{87}$Rb and $^{133}$Cs respectively. The mixture contains  $3.0(3)\times10^{5}$ $^{87}$Rb atoms, and $2.6(4)\times10^{4}$ $^{133}$Cs atoms at $0.32(1)~\mu$K. This temperature is well below the \emph{p}-wave threshold ($k_{\rm{B}}\times$~56~$\mu$K based upon the published $C_{6}$ coefficient \cite{Marinescu1999}), simplifying the interpretation of the Feshbach spectrum. For each experimental cycle the ultracold mixture is allowed to evolve at a specific homogeneous magnetic field for 5~s, and then the $^{133}$Cs atom number is measured. In Fig. \ref{fig:RbCs} the detection of a Feshbach resonance near $180$~G is presented, each data point corresponds to an average of 3-5 measurements.

By fitting a Lorentzian function to the data shown in Fig. \ref{fig:RbCs} the position of the Feshbach resonance was measured to be $B_{0} =181.7(5)$~G. This result is in excellent agreement with the previous measurement of this resonance \cite{Pilch2009}. Using this setup we are currently extending the interspecies Feshbach resonance search up to bias fields in excess of 1150~G, and to $^{85}$Rb and $^{133}$Cs mixtures. Locating an interspecies Feshbach resonance now opens the possibility of creating RbCs molecules via magneto-association.

\section{\label{chap:Molecules}Feshbach Association of Cs$_{2}$ Dimers}

The second step towards the production of ground state molecules requires the controlled magneto-association of weakly bound molecules using a Feshbach resonance. To develop experimental protocols and detection methods the magneto-association of Cs$_{2}$ dimers has been explored using a well characterised $^{133}$Cs Feshbach resonance \cite{Herbig2003}. To create a high phase-space density sample of only $^{133}$Cs, both $^{87}$Rb and $^{133}$Cs are loaded into the magnetic quadrupole trap and the standard sympathetic cooling of $^{133}$Cs via $^{87}$Rb is performed (see Fig. \ref{fig:cooling}). However the final RF frequency is decreased to cut away all of the $^{87}$Rb atoms during the loading of the dipole trap. Further evaporation is performed until the $^{133}$Cs cloud is close to degeneracy.

The $|3, +3 \rangle$ state used has a 4(g)4 Feshbach resonance at $19.8$~G \cite{Vuleti1999} that is estimated to be $5$~mG wide \cite{Herbig2003}. (The 4(g)4 notation refers to the $F=4, l=4$ and $m_F=4$ molecular state, where \emph{F} is the total internal angular momentum, \emph{l} is the rotational angular momentum, and $m_{F}$  is the projection along the quantization axis.) This resonance is crossed from high to low field at a rate of $47$~G~s$^{-1}$. The field is then rapidly jumped to $17$~G to null atomic interactions and the dipole trap is switched off. Stern-Gerlach separation is immediately applied to spatially distinguish between the atomic and molecular samples. The magnetic field gradient is fixed at 31.1~G~cm$^{-1}$ to levitate the atomic cloud. During this variable hold time the molecules fall down away from the levitated atoms due to their smaller magnetic moment to mass ratio. To dissociate the molecules for imaging the magnetic field is jumped back across the resonance to 21~G in $130~\mu$s. This reverse sweep brings the molecules above the scattering continuum, from here they rapidly dissociate into free atoms. Finally the magnetic field gradient is switched off for a 2~ms free expansion before an absorption image reveals the distribution of the dissociated molecules (and the free atoms).

\begin{figure}[h t]
\centering
\includegraphics[angle=0, trim = 13mm 13mm 13mm 13mm, clip, scale=0.323]{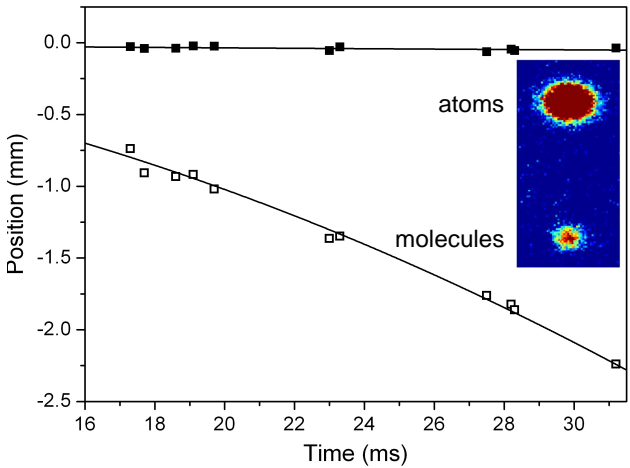}
\caption{The production of Cs$_{2}$ molecules using a Feshbach resonance. Open (closed) symbols mark the molecular (atomic) clouds position as a function of the levitation time. From these data a molecular acceleration of 3.86(4)~m/s$^{2}$ is measured, this corresponds to $\mu = 0.92(1) \mu_{B}$. Inset: A typical absorption image taken after the molecules have been dissociated back into atoms. At the top the levitated atomic cloud contains $1.0\times10^{5}$ atoms, the smaller molecular sample below contains $7.5\times10^{3}$ Cs$_{2}$ Feshbach molecules.}
\label{fig:Cs2}
\end{figure}

The atomic and molecular clouds' positions as a function of the levitation time is presented in Fig. \ref{fig:Cs2}. The inset to this figure shows a typical absorption image; at the top a large atom cloud is levitated while a smaller molecular cloud falls away. Typical magneto-association efficiencies are $\sim$12\% and samples containing 7,000 molecules are produced from the high phase-space density $^{133}$Cs cloud. The molecular acceleration can be measured from the data presented in Fig. \ref{fig:Cs2} and this allows the magnetic moment for the Cs$_{2}$ molecules to be calculated. The molecular acceleration at 31.1~G~cm$^{-1}$ is measured to be 3.86(4)~m~s$^{-2}$, this corresponds to a magnetic moment $\mu = 0.92(1)\mu_{\rm{B}}$. This value concurs with another measurement performed by levitating the molecular cloud at 51.4(2)~G~cm$^{-1}$. These results are in good agreement with previous work and theoretical calculations \cite{Herbig2003}.

\section{\label{chap:Conclusion}Conclusion and Outlook}

In this paper we have presented a simple levitated crossed dipole trap which is easily loaded from a magnetic quadrupole trap and is suitable for the sympathetic cooling of $^{133}$Cs by $^{87}$Rb to high phase-space densities. The simplicity of the setup and method is in contrast to previous work on ultracold mixtures of $^{87}$Rb and $^{133}$Cs \cite{Pilch2009}. We believe our approach could be readily implemented into many existing experimental setups that have Helmholtz and anti-Helmholtz coils already in position.

Using this simple approach we have succeeded in making a high phase-space density mixture of $^{87}$Rb and $^{133}$Cs combatting strong three-body losses en route through careful optimisation of the evaporative cooling. This high phase-space density mixture has the appropriate starting conditions for efficient magneto-association. With this in mind we have observed an interspecies Feshbach resonance at 181.7(5)~G and have created Cs$_{2}$ dimers using a $^{133}$Cs Feshbach resonance to test the experimental protocol.

The next step is to perform Feshbach association of $^{87}$Rb and $^{133}$Cs into weakly bound heteronuclear molecules via an interspecies Feshbach resonance. A broad entrance-channel dominated resonance such as that shown in Fig. \ref{fig:RbCs} could be an ideal candidate for Feshbach association \cite{Kohler2006,Chin2010}. Exploration of the bound state spectrum close to threshold \cite{Mark2007} should then allow the identification of a molecular state suitable for optical transfer to the rovibrational ground state. We are currently developing the STIRAP laser system, following the transfer scheme first proposed in ref. \cite{Stwalley2004}. The optical excitation of the Feshbach molecules is at a wavelength of $\sim1550$~nm. The wavelength of the second photon is then selected to remove the $3836.14(50)$~cm$^{-1}$ of binding energy \cite{Piotr2010} of the rovibrational molecular ground state. We expect ultracold ground state RbCs molecules to be within reach shortly in our experiments.

We acknowledge financial support from the EPSRC (GR/S78339/01, EP/E041604/1, EP/H03363/1) and the European Science Foundation within the framework of the EuroQUAM collaborative research project QuDipMol. SLC acknowledges the support of the Royal Society.


\end{document}